\documentclass[11pt]{article}
\usepackage{moriond,epsfig}

\def\ra{\rightarrow}

\def\be{\begin{equation}}
\def\ee{\end{equation}}
\def\bea{\begin{eqnarray}}
\def\eea{\end{eqnarray}}

\def\kpme{K^+\rightarrow\pi^+\mu^+ e^-}
\def\pke3{K^+ \rightarrow \pi^0 e^+ \nu}
\def\ake3{K \rightarrow \pi  e \nu}
\def\ke3g{K^+\rightarrow\pi^0 e^+ \nu \gamma}
\def\kmu3{K^+\rightarrow\pi^0\mu^+ \nu}

\def\ke4{K^+\rightarrow\pi^+\pi^-e^+\nu}
\def\kpipi{K^+\ra \pi^+\pi^0}

\def\k3pi{K^+\ra \pi^+ \pi^0 \pi^0}
\def\ktau{K^+\ra \pi^+ \pi^+ \pi^-}

\def\dal{\pi^0 \ra e^+ e^- \gamma}

\def\Mnu2{M_{\nu}^{2}}

\begin{document}

\vspace*{4cm}
\title{
\boldmath $\pke3$  BRANCHING RATIO FROM E865}

\author{ A.A. POBLAGUEV~\footnote{
Representing the BNL E865-Collaboration: 
D.M.~Lazarus, H.~Ma, P.~Rehak (Brookhaven National laboratory);
G.S.~Atoian, V.V.~Issakov, A.A.~Poblaguev (INR, Moscow);
J.~Egger, W.~Harold, H.~Kaspar (Paul Scherrer Institut);
B.~Bassalleck, S.~Eilerts, H.~Fischer, J.~Lowe (New Mexico);
R.~Appel, N.~Cheung, D.E.~Kraus, P.~Lichard,  A.~Sher, J.A.~Thompson 
(Pittsburgh);
D.R.~Bergman, S.~Dhawan, H.~Do, J.~Lozano, W.~Majid, M.E.~Zeller (Yale);
S.~Pislak, Aleksey Sher, P.~Tru\"{o}l (Z\"{u}rich)
} }

\address{
Institute for Nuclear Research of Russian Academy of Sciences, Moscow 117312, Russia\\
and \\
Physics Department, Yale University, New Haven, CT 06511, USA
}

\maketitle\abstracts{
E865 at the Brookhaven National Laboratory AGS collected about 70,000
$K^+_{e3}$
events to measure the $K^+_{e3}$ branching ratio
relative to the observed $\kpipi$, $\kmu3$, and $\k3pi$ decays. 
The $\pi^0$ in all the decays was detected using the $e^+ e^-$ pair 
from $\dal$ decay and no photons were required.
Using the Particle Data Group branching ratios 
\cite{pdg} for the normalization decays we obtain 
$BR(K^{+}_{e3(\gamma)})=(5.13\pm0.02_{stat}\pm0.09_{sys}\pm0.04_{norm})\%$, 
where $K^{+}_{e3(\gamma)}$ includes the effect of virtual and real photons.
This result is $\approx 2.3$ standard deviations higher than the current 
Particle Data Group value. 
Implications  for the $V_{us}$ element of the CKM matrix, and the matrix's
unitarity are discussed.
}

 The experimentally determined  Cabibbo-Kobayashi-Maskawa (CKM) matrix
describes quark mixing in
the standard model framework.
Any deviation from the matrix's unitarity would undermine 
the validity of the standard model.
One unitarity condition involves the first row 
elements:
\begin{equation}
|V_{ud}|^2+|V_{us}|^2+|V_{ub}|^2=1-\delta
\label{unitest1}
\end{equation}
where a non-zero value of $\delta$ indicates a deviation from unitarity.
The $V_{ud}$ element is obtained from nuclear and neutron decays.
$V_{ub}$, from 
the semileptonic decays of B mesons \cite{pdg}, is too small to 
affect  Eq. (\ref{unitest1}).
The $V_{us}$ element can be determined either from hyperon, 
$K\rightarrow \pi \mu \nu$($K_{\mu 3}$) or from $\ake3$ ($K_{e3}$)
decays. However, $K_{e3}$ decays provide a smaller 
theoretical
uncertainty \cite{pdg,lut}. The most precise value of 
$V_{ud}$ obtained from
the nuclear superallowed Fermi beta decays leads to 
$\delta=(3.2\pm1.4)\cdot 10^{-3}$  \cite{hardy},
 a 2.3$\sigma$ deviation from unitarity.

Both experimental and theoretical efforts to improve
the determination of $V_{ud}$ continue.
Theoretical contributions to $V_{us}$  were reevaluated
recently \cite{rad,arie,calderon,bijnens}, 
but there has been little new experimental input on 
the $K^+_{e3}$ branching ratio. Since
the $|V_{ud}|^2$ and $|V_{us}|^2$  uncertainties are comparable,
a high statistics measurement of the $K^+_{e3}$
branching ratio (BR) with good
control of systematic errors is useful.

The bare (without QED corrections) 
$K^+_{e3}$ decay rate \cite{lut,rad,arie,dafne} is:
\begin{equation}
d\Gamma(K^{+}_{e3})={C(t)|V_{us}|^{2}}
|f_{+}(0)|^2[1+\lambda_{+}\frac{t}{M_{\pi}^2}]^{2}
dt
\label{ke3_rate}
\end{equation}
where 
$t=(P_{K}-P_{\pi})^2$,
C(t) is a known kinematic function, 
and $f_{+}(0)$ is the vector form factor value
at $t=0$,  determined theoretically \cite{lut,rad}.
Two recent 
experiments \cite{kek,ISTRA+} give $\lambda_+$ (the form factor slope)
measurements consistent with each other and with previous measurements.
An omitted negligible term contributing to Eq. \ref{ke3_rate} contains the form
factor $f_{-}$, and is proportional to $M_{e}^{2}/M_{\pi}^{2}$.
With current constraints on the hypothetical tensor and
scalar form factors \cite{ISTRA+}, their possible contribution 
to the $K^+_{e3}$ decay rate are also neglibible.

\begin{figure}[hbt]
\setlength \epsfysize{14cm}
\setlength \epsfxsize{7cm}
\begin{minipage}[t]{0.47\textwidth}
\centering{\includegraphics[scale=0.28]{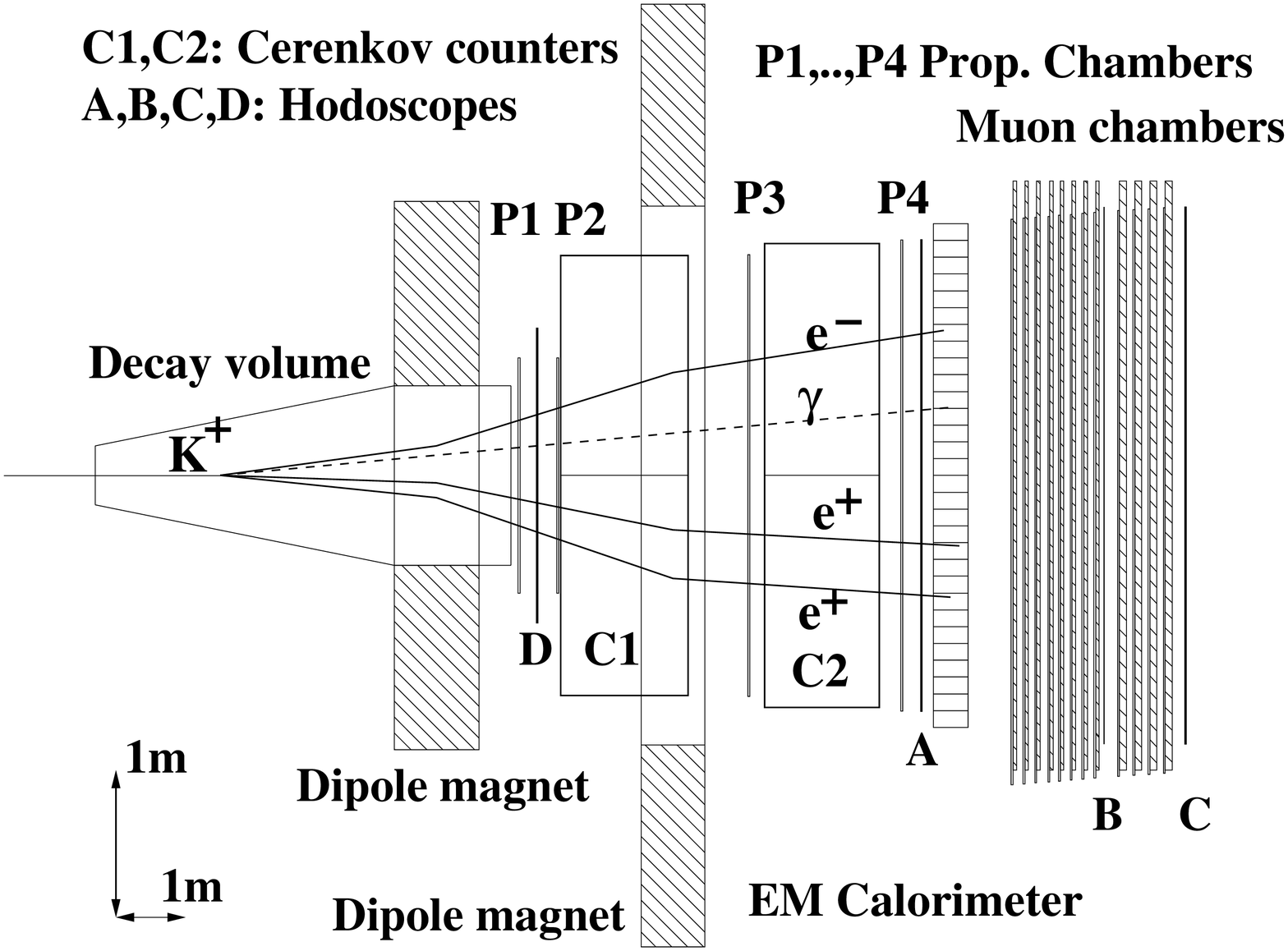}}
\caption{Plan view of the E865 detector with a simulated $\pke3$ decay
followed by $\dal$.
\label{e865_det}
}
\end{minipage} \hspace{0.05\textwidth} \begin{minipage}[t]{0.47\textwidth}
\centering{\raisebox{0.9cm}{\includegraphics*[scale=0.415]{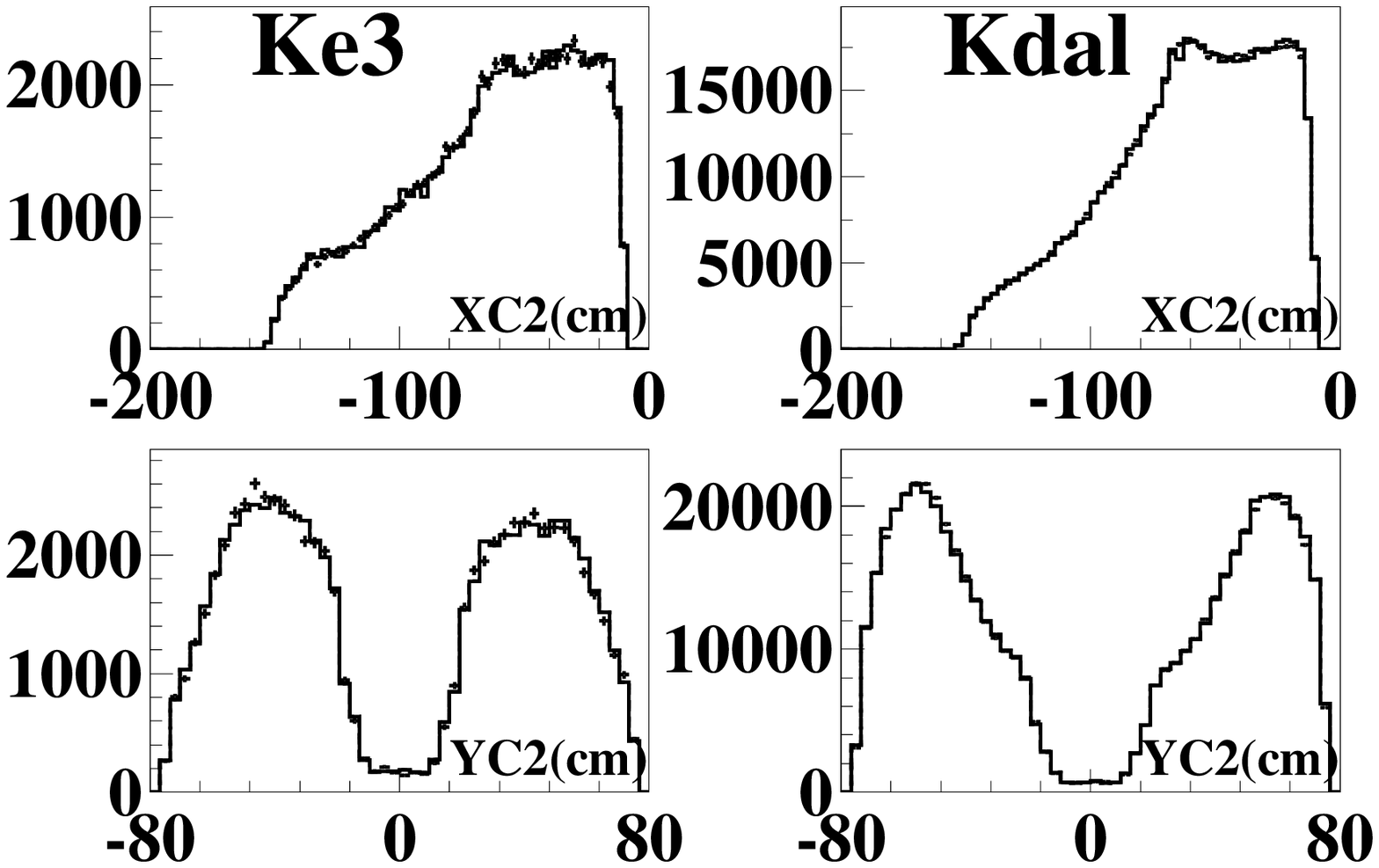}}}
\caption{Distributions of X and Y positions of the first positive track
(not $e^+$ from the $\pi^0$ decay)
 for the selected $K^+_{e3}$ and {\sf Kdal} samples.
X and Y positions are measured at the end of the second pair of 
the \v{C}erenkov counters (C2). Histograms represent Monte Carlo;
points with errors represent data.}
\label{xyc23_mee1}
\end{minipage}
\end{figure}

E865 \cite{nim} searched for the lepton flavor 
violating decay $\kpme$. The detector (Fig. \ref{e865_det}) 
resided in a 6 GeV/c positive beam \cite{nim}. For the $K^{+}_{e3}$ running \cite{PRL}, 
the
intensity was reduced by a factor of 10, to $10^7$ kaons, 
$2 \times 10^8$ protons, and $2 \times 10^8 \pi$ per $2.8\ \mathrm{s}$ spill.
The beam was intentionally debunched at extraction to remove rf
structure at the experiment.
The first
dipole magnet separated particles by charge, while the second magnet
together with four multiwire proportional chambers (MWPCs: P1-P4)
formed the spectrometer. The particle identification used  the
 threshold multichannel \v{C}erenkov detectors (C1 and C2, each separated 
into left and right volumes, for four independent counters) filled with
gaseous methane
(\v{C}erenkov threshold $\gamma_t \approx 30$  and  electron
detection efficiency $\epsilon_{e} \approx 0.98$ 
\cite{ke3_th}),
an electromagnetic calorimeter \cite{nim}, and a muon detector
(not used for the $K^{+}_{e3}$ measurement). The D
and A 
scintillator
hodoscopes gave left/right and crude vertical position. 

The $\pi^0$ from the kaon decays
was detected through the $e^+ e^-$ from the $\dal$ decay, 
with the $\gamma$
detected in some cases.
To eliminate the uncertainty (2.7\%) of the 
$\dal$ BR,
and to reduce  systematic uncertainty, we used the other three 
major decay modes with a $\pi^0$
in the final state ($\kpipi$($K^{+}_{\pi2}$), $K^{+}_{\mu 3}$, 
$\k3pi$($K^{+}_{\pi3}$)) 
for the  normalization sample ({\sf Kdal}).

The $K^+_{e3}$ data was collected in a
one-week dedicated run in 1998, with special on-line trigger logic.

The   {\sf Kdal} and  
$K^+_{e3}$ data  were collected by the $\mathsf{e^{+}e^{-}}$ trigger, which
was designed to detect $e^{+}e^{-}$ pairs and required at least one 
D-counter scintillator slat on each (left and right) side of the detector
and signals from each of  the four \v{C}erenkov counters.
The 
\v{C}erenkov efficiency trigger 
required only 3 out of 
4 \v{C}erenkov counters (no D-counter requirement).
The  {\sf TAU} trigger, requiring only two D-counter scintillator
hits (one left, and one right), collected events for the $\ktau$ ($K_{\tau}$)
 sample, to  study of the detector unbiased by 
\v{C}erenkov requirements.
About  $50\times10^6$ triggers were accumulated,
 $\approx$ $37\times10^6$  in the $\mathsf{e^{+}e^{-}}$ trigger.
About 75\% of $\mathsf{e^{+}e^{-}}$ triggers included accidental tracks,
often a $\mu$ from high momentum 
$K\rightarrow \mu\nu$ or $\pi\rightarrow \mu\nu$ decays partially satisfying 
the
\v{C}erenkov requirement.

Off-line  reconstruction used  the spectrometer
only.
The \v{C}erenkov and D counter efficiencies were obtained from the 
\v{C}erenkov efficiency triggers.
 The redundancy of the MWPCs (4 planes/chamber)
and track reconstruction  was used to extract MWPC efficiencies.
The absence of the electromagnetic calorimeter from the trigger allowed 
its efficiency determination.
Each  efficiency was measured over its  relevant phase space.

Relevant kaon decay chains \cite{ke3_th} 
were simulated with  GEANT \cite{geant} (including decays 
of secondary pions and muons).
For $K^{+}_{e3}$,  $\lambda^+ = 0.0278 \pm 0.0019$ \cite{pdg} was used.
The radiative corrections to the $K^+_{e3}$ decay phase-space 
density \cite{rad}
were used.
The $K^+_{e3\gamma}$ (inner bremsstrahlung) decays outside the 
$K^+_{e3}$ Dalitz plot boundary were explicitly simulated 
\cite{dafne}. 
For $\dal$ decay, radiative corrections were taken into account
according to Ref. \cite{mik}.
  Measured efficiencies were 
applied \cite{ke3_th}, and accidental detector hits (from reconstructed 
$K_{\tau}$ events) were added.
About 10$\%$ of both the $K^{+}_{e3}$ and {\sf Kdal} samples had extra reconstructed
tracks.

Selection criteria, common to $K^+_{e3}$ and {\sf Kdal}, included 
requirements for a good quality three track  vertex in the decay
volume (no requirement for exactly three reconstructed tracks was applied), 
for the
three tracks to cross the active parts of the detector,  for the low
($M_{ee}<0.05$ GeV) mass $e^+e^-$ pair to be identified in 
the \v{C}erenkov counters, and for the second positive track to have less 
than 3.4 GeV/c momentum.  The momentum cut rejects events where 
$\mu^{+}$ or $\pi^{+}$ from {\sf Kdal} decays is above \v{C}erenkov threshold
and can be identified as $e^+$.
  A geometric \v{C}erenkov
ambiguity
cut   rejected events
(27$\%$, 15$\%$, 25$\%$, and 35$\%$  for $K^{+}_{e3}$, $K^{+}_{\pi2}$, $K^{+}_{\mu3}$,
and $K^{+}_{\pi3}$, respectively) where the \v{C}erenkov counter
response could not be
unambiguously assigned to separate tracks \cite{ke3_th}.

The $K^+_{e3}$ sample was then selected by requiring the second positive 
track to 
be identified as $e^+$ in 2 of the 3  electron detectors: C1, C2, or the
calorimeter, each with $\epsilon_{e} \approx 98\%$.
Events entering the {\sf Kdal} sample
had no response in at least one of the two \v{C}erenkov counters.
These criteria minimized systematic uncertainties \cite{ke3_th}, but resulted
in a small overlap, $\approx 3\%$ of the  $K^+_{e3}$ sample and
$\approx 0.3\%$ of the {\sf Kdal} which 
was accounted for in the BR  calculation.
The $K^{+}_{\pi 2}$ acceptance is $\approx 1.2 \%$.
The $K^+_{e3}$ acceptance  $\approx0.7\%$ \cite{ke3_th}, somewhat lower
because of the lower average $e^+$ momentum in the $K^+_{e3}$ decay.
The acceptance can be approximately understood by assuming a factor of three
loss for each charged particle, 30 $\%$ for the \v{C}erenkov ambiguity,
and approximately a factor of 2 for other cuts.
Final acceptances for the
three modes in the {\sf Kdal}  sample
differed by $\le  4\%$ taking into account that
either of the $\pi^0$s from $K^{+}_{\pi 3}$ can decay into $e^+e^-\gamma$.
The final $K^{+}_{e3}$ and {\sf Kdal} samples were
$71\,204$ and $558\,186$, respectively. Figure \ref{xyc23_mee1} shows 
 some relevant spatial distributions.

\begin{figure}[hbt]
\setlength \epsfysize{8cm}
\setlength \epsfxsize{10cm}
\begin{minipage}[t]{0.47\textwidth}
\centering{\includegraphics[scale=0.30]{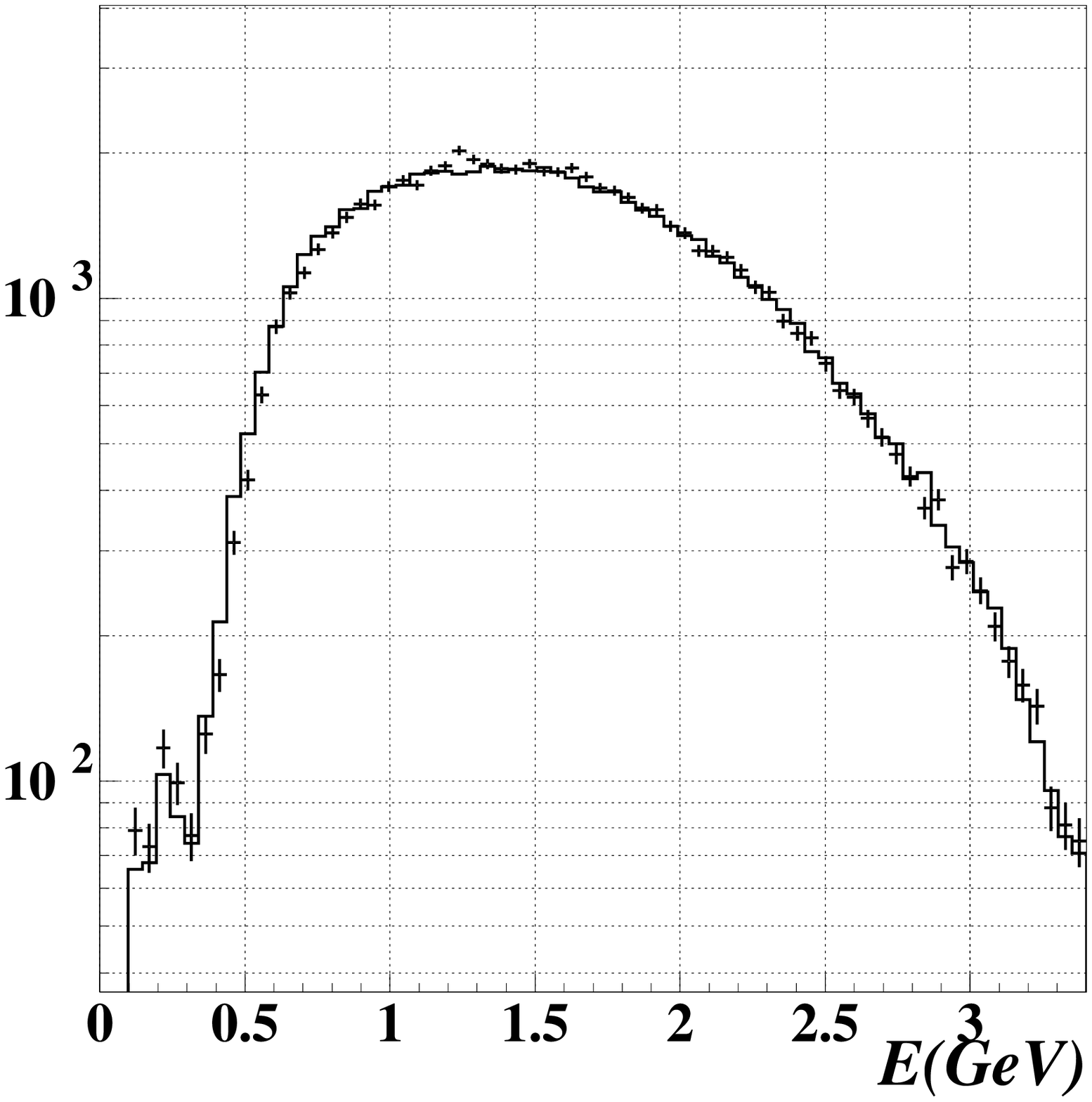}}
\caption{
Energy deposited in the calorimeter by the second positive track from 
the selected $K^+_{e3}$ sample ($e^+$ which is not from the
low mass $e^+e^-$ pair).
No calorimeter information was used for the $e^+$ identification.
Markers with errors represent data; the histogram is simulation.
\label{e3_ke3_22}
}
\end{minipage} \hspace{0.05\textwidth} \begin{minipage}[t]{0.47\textwidth}
\centering{\raisebox{0.5cm}{\includegraphics*[scale=0.415]{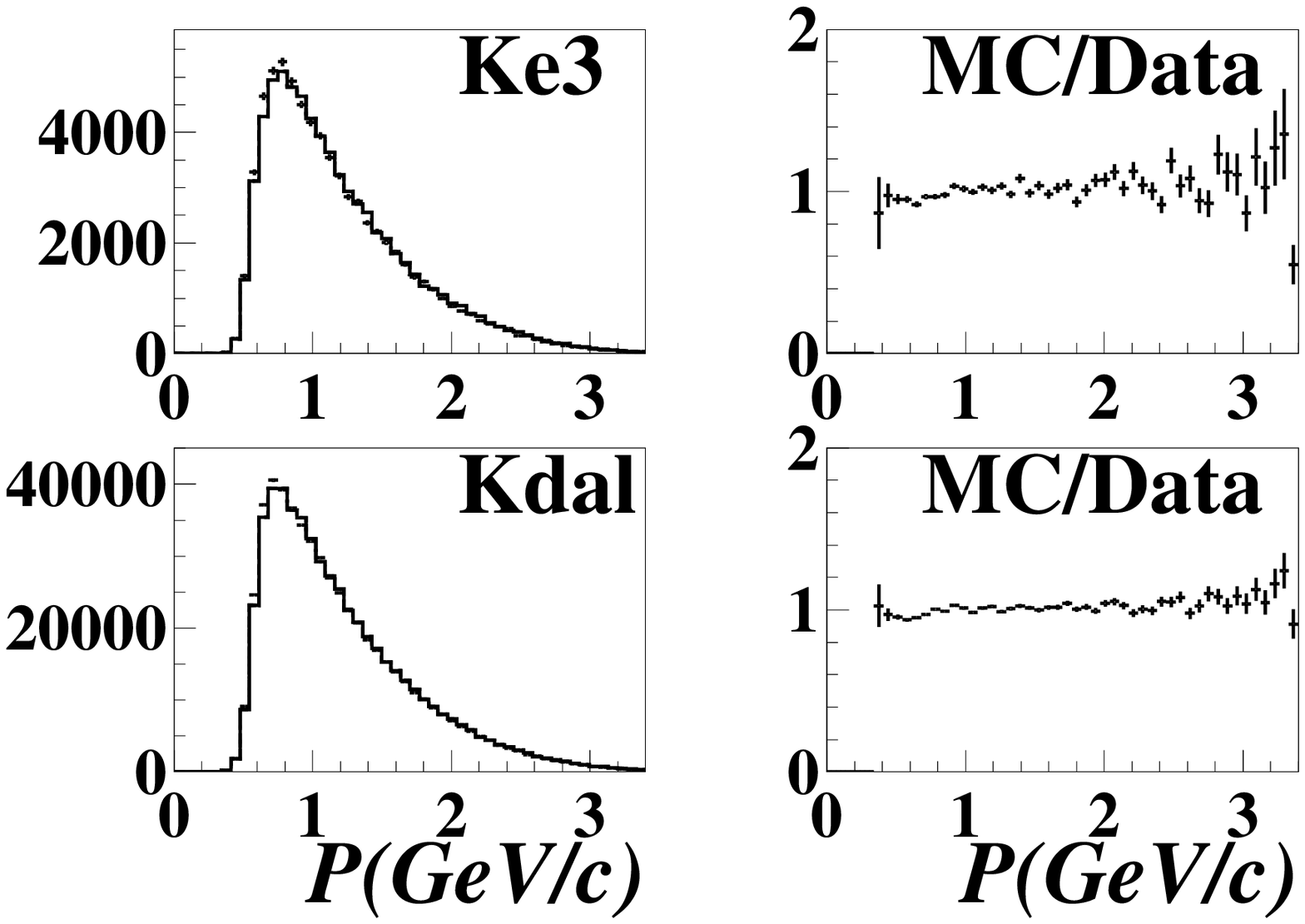}}}
\caption{Reconstructed momentum of the $e^+$ from
the low mass $e^+ e^-$ pair from the selected $K^+_{e3}$ and 
{\sf Kdal} samples.
Histograms represent Monte Carlo;
points with errors represent data. Plots on the right show the bin by bin 
Monte Carlo to data ratio.}
\label{p12_ke3}
\end{minipage}
\end{figure}

Contamination of the $K^+_{e3}$ sample by other $K^+$ decays
occurred when  $\pi^+$ or
$\mu^+$ from {\sf Kdal} decays were misidentified as $e^+$, or 
as a result of $\pi^0\rightarrow e^+ e^- e^+ e^-$.
Contamination due to secondary particle decays
was estimated to be at the level of 0.1\%.
About 8\% of final state pions decayed into muons inside the spectrometer.
The careful MWPC simulation gave good agreement of reconstructed 
track $\chi^2$ and vertex distributions between data and Monte Carlo 
calculation.
No tight track $\chi^2$ cuts were applied, and the
systematic uncertainties estimated by variation of the
vertex cuts were included in the final result.
The check of  BR( $K_{\tau}$/{\sf Kdal}),
described below, also tests the
 final state $\pi$ and $\mu$ decays.

Total contamination of the Ke3 sample
 was estimated to be $(2.49\pm0.05_{stat}\pm0.32_{sys})\%$, with
the systematic uncertainty 
caused by 
the simulation accuracy of 
the C1 and C2 response to $\pi^+$ and $\mu^+$.
Contamination due to overlapping events was $(0.25\pm0.07)\%$ and 
$(0.12\pm0.05)\%$ of the {\sf Kdal}  and $K^{+}_{e3}$,
 respectively.
Figure \ref{e3_ke3_22} shows the energy 
distribution 
in the calorimeter from the
$e^+$ in the $K^+_{e3}$ sample. The contamination
is manifest in the minimum ionization spike at 250 MeV.
The small excess of  data
in the spike
agrees with our contamination uncertainty estimate.

The final $K^+_{e3}$ sample included $\approx$30\% of events with 
a fully reconstructed $\pi^0$. We used the $\pi^0$ information as a
consistency
check. Not requiring $\pi^0$  in our main analysis
 minimized the uncertainty arising from
photon detection and reconstruction in the calorimeter, but 
 increased vulnerability
to contamination from upstream decays and photon conversion.
Upstream decays whose photon produced pairs
before the decay volume (evacuated to about 
$10^{-8}$ nuclear interaction length) were suppressed by requiring
the three track vertex to be more than $2\ \mathrm{m}$ downstream of
the decay volume entrance. In addition, 
the results obtained from the two independent
samples, one with  and one without the $\pi^0$ reconstructed,
did not show a statistically significant discrepancy.

The $K^+_{e3}$ 
 statistical precision is $0.4\%$.
The systematic error estimate,   summarized in Table \ref{tab:sys}, 
was
determined from the BR  stability under
variation of reconstruction procedure, 
selection criteria, assumed detector efficiencies,
 and subdivision of 
both $K^{+}_{e3}$ and {\sf Kdal} samples \cite{ke3_th}. No significant correlations
between any of the 
different systematic uncertainties were observed.

\begin{table}[hbt]
\begin{minipage}[c]{0.66\textwidth}
\begin{tabular}{|l|p{50pt}|}
\hline
 Source of systematic error &
Error estimate\\
\hline
 Magnetic field uncertainty          & 0.3$\%$ \\
\hline

 Vertex finding and quality cut	     & 0.6$\%$\\
\hline


 Vertex position cut			  & $0.4\%$\\
\hline

 \v{C}erenkov Ambiguity Cut		  & $0.3\%$ \\
\hline

  $M_{ee}$ cut                      & $0.2\%$ \\
\hline

 Detector Aperture		  & $0.2\%$ \\
\hline

 $(\pi/\mu)^+$ identification      & $0.04\%$ \\
\hline

 MWPC efficiencies		  & $0.2\%$ \\
\hline

 D counter efficiencies		  & $ 0.15\%$\\
\hline

 \v{C}erenkov efficiencies		  & $0.3\%$ \\
\hline

 Contamination of the selected samples  & $0.3\%$  \\
\hline
Removal of extra tracks         & $0.2\%$   \\
\hline

 Vertical spatial/angle distributions discrepancy &  
                                   $0.8\%$  \\
\hline
    
 $e^+/e^-$ momentum distributions discrepancy
                          	  & $1.3\%$  \\
\hline

 $K^{+}_{e3}$ trigger efficiency
                              	  & $0.1\%$  \\
\hline

 Uncertainty in the $K^+_{e3}$ form factor slope
                              	  & $0.1\%$  \\
\hline

\hline

Total error		           &  $1.8\%$ \\
\hline
\end{tabular}
\end{minipage} \hspace{0.01\textwidth} 
\begin{minipage}[t]{0.33\textwidth}
\vspace*{-5.1cm}\caption{Systematic uncertainty sources and estimates of their respective
contributions  to the final result's uncertainty. The total
systematic error is 
the sum (in quadrature) of the  individual contributions.
\label{tab:sys}
}
\end{minipage}
\end{table}

The two largest contributions to the systematic  error  come from
the discrepancies 
\cite{ke3_th} between data and Monte Carlo in the momentum 
(Fig. \ref{p12_ke3}) and spatial distributions.
These
errors were determined by dividing $K^+_{e3}$ and {\sf Kdal}
events into two roughly equal samples, using the relevant parameters, 
and observing the variation of the result \cite{ke3_th}. The errors were
found to be uncorrelated.
The sensitivity of the vertical spatial discrepancy to 
the MWPC alignment and of the momentum discrepancy to the spectrometer 
parameters
is indicative of their possible origins \cite{ke3_th}.
The Z-vertex position is also sensitive to the magnetic field, but has a 
smaller systematic error contribution as determined from both upstream and
downstream cuts in Z.

As an additional consistency check, we estimated the $K_{\tau}$/{\sf Kdal}
BR.  The result
was $(1.01\pm 0.02)\times$ the Particle Data Group (PDG) ratio \cite{pdg},
(the theoretical prediction \cite{dalitz} was used 
for the $\dal$ decay rate). The 2$\%$ error  was 
dominated by the uncertainty
in the prescale factor of the {\sf TAU} trigger.
A second consistency check compared the $K^+_{e3}$ BR
from 1998 and  1997 data.
The 1997 $K^+_{e3}$ data used a trigger that required
calorimeter hits, and  A and D-counters. That trigger 
neither allowed measurement of these detector efficiencies
nor of the trigger 
efficiency.
While we did not use the 1997 data for our final result, 
the 1997 $K^+_{e3}$ branching ratio was statistically consistent 
(within one sigma) with that from  1998.
This agreement is important since the momentum spectrum
discrepancy between data and Monte Carlo in the
1997 data is qualitatively different from 1998 \cite{ke3_th}.
A preliminary reconstruction version was used for the 1997 data, without
the final magnetic field and detector alignment. This bolsters our 
intuition that the discrepancies in decay product momenta and spatial
distributions, which dominate the systematic uncertainties, reflect 
our imperfect knowledge of the magnetic field and detector positions but 
do not bias our result beyond our estimated systematic errors. 

We estimated the form factor slope $\lambda_+$ from both 1998 and 1997
$K^+_{e3}$ data \cite{ke3_th}. We
obtained: $\lambda_{+}=0.0324\pm0.0044_{stat}$ for 1998,
and $\lambda_{+}=0.0290\pm0.0044_{stat}$ for the 1997 data, both 
consistent with the current PDG fit.

After contamination subtraction \cite{ke3_th},
our result is $BR(K^{+}_{e3(\gamma)})/(BR(K^{+}_{\pi 2}+K^{+}_{\mu 3}+
K^{+}_{\pi 3}))
=0.1962\pm0.0008_{stat}\pm0.0035_{sys}$,
where $K^+_{e3(\gamma)}$ 
includes all QED contributions (loops and inner bremsstrahlung).
As noted above, the $\pi^0$ was detected using 
the $e^+ e^-$ pair from $\dal$  and no photons were required.

Using current \cite{pdg}  {\sf Kdal} BR's
we infer 
$BR(K^{+}_{e3(\gamma)})=(5.13\pm0.02_{stat}\pm0.09_{sys}\pm0.04_{norm})\%$,
where the normalization error was determined by the PDG estimate
of the {\sf Kdal}  BR  uncertainties. This result does not include
the correction due to the correlation of the PDG kaon decay ratios, since
it was estimated to be small compared to the systematic
error.
The PDG fit to the previous $K^+$ decay 
experiments yields $BR(\pke3)=(4.87\pm0.06)\%$ \cite{pdg},
$\approx 2.3$ standard deviations lower than our result.

Radiative corrections for decays inside the $K^{+}_{e3}$ Dalitz plot boundary
were estimated to be $-$1.3\% using the procedure of Ref. \cite{rad};
 $K^{+}_{e3\gamma}$ decays outside the Dalitz 
plot boundary gave  +0.5\%. Thus, the total radiative correction
was $-0.8\%$
resulting in the bare 
$BR(K^{+}_{e3})=(5.17\pm0.02_{stat}\pm0.09_{sys}\pm0.04_{norm})\%$.

Using the  PDG value for $G_{F}$, 
the short-distance enhancement factor 
$S_{EW}(M_{\rho},M_{Z})=1.0232$ \cite{rad,sirnew},
and our result for the bare $K^+_{e3}$  rate we obtain 
$|V_{us}f_{+}(0)|=0.2243 \pm 0.0022_{rate} \pm 0.0007_{\lambda_{+}}$, 
which gives
$|V_{us}|=0.2272 \pm0.0023_{rate} \pm 0.0007_{\lambda_{+}} \pm 0.0018_{f_{+}(0)}$
if $f_{+}(0)=0.9874\pm0.0084$ \cite{lut,rad}. With this value of $V_{us}$
and $V_{ud}$ from superallowed nuclear Fermi beta decays \cite{hardy},
$\delta=-0.0003 \pm 0.0016$.

This result is consistent with CKM unitarity, and is in agreement with recent theoretical determinations of $V_{us}$ from hyperon decays, $0.2250\pm0.0027$ \cite{cabibbo}, and from lattice calculations of pseudoscalar decay constants, $0.2236\pm0.0030$ \cite{marciano}.
However, it  increases the discrepancy 
with the value of $V_{us}=0.2176\pm0.0025$ from $K^0_{e3}$ decay if extracted
under conventional theoretical assumptions about symmetry breaking
\cite{0401173,KLOE} \footnote{%
The recent preliminary value of $V_{ud}$ from
the $K^0_S\to\pi e\nu$ presented by the KLOE Callaboration at this
Conference \cite{KLOE-new} shows much better consistency with our
result.}. 
The KLOE $K_{e3}$  measurements for neutral and charged kaons currently 
in progress \cite{KLOE,KLOE-new}
should help to clarify the experimental situation.

We thank V. Cirigliano for the $K^+_{e3}$ radiative corrections code.
We gratefully acknowledge the contributions  by the staffs  of the AGS,
and   participating institutions. This work was supported in part by 
the U.S. Department of Energy under contract DE-AC02-98CH10886, 
the National Science Foundations of 
the USA, Russia and Switzerland, and the Research Corporation.

\end{document}